\documentclass[useAMS,usenatbib]{mn2e}
\usepackage{fourier}
\usepackage{graphicx}

\title[The outflow in G23.44-0.18]{Outflow activities in the young high-mass stellar object G23.44-0.18}
\author[Ren et al.]{Jeremy Zhiyuan Ren$^{1}$, Tie Liu$^{1}$, Yuefang Wu$^{1}$, and Lixin Li$^{1,2}$ \\
$^{1}${Department of Astronomy, Peking University, 100871, Beijing China, E-mail: rzy,ywu@pku.edu.cn} \\
$^{2}${The Kavli Institute for Astronomy and Astrophysics, Peking University, Yi He Yuan Lu 5, Hai Dian Qu, Beijing 100871, P. R. China}}
\begin{document}

\pagerange{\pageref{firstpage}--\pageref{lastpage}} \pubyear{2011}

\maketitle

\label{firstpage}

\begin{abstract}
We present an observational study towards the young high-mass star forming region G23.44-0.18 using the Submillimeter Array. Two massive, radio-quiet dusty cores MM1 and MM2 are observed in 1.3 mm continuum emission and dense molecular gas tracers including thermal CH$_3$OH, CH$_3$CN, HNCO, SO, and OCS lines. The $^{12}$CO (2--1) line reveals a strong bipolar outflow originated from MM2. The outflow consists of a low-velocity component with wide-angle quasi-parabolic shape and a more compact and collimated high-velocity component. The overall geometry resembles the outflow system observed in the low-mass protostar which has a jet-driven fast flow and entrained gas shell. The outflow has a dynamical age of $6\times10^3$ years and a mass ejection rate $\sim10^{-3}~M_{\odot}$ year$^{-1}$. A prominent shock emission in the outflow is observed in SO and OCS, and also detected in CH$_3$OH and HNCO. We investigated the chemistry of MM1, MM2 and the shocked region. The dense core MM2 have molecular abundances of 3 to 4 times higher than those in MM1. The abundance excess, we suggest, can be a net effect of the stellar evolution and embedded shocks in MM2 that calls for further inspection.
\end{abstract}


\begin{keywords}
stars: pre-main sequence --- ISM: molecules --- ISM: abundances --- ISM: kinematics and dynamics --- ISM: individual (G23.44-0.18) --- stars: formation
\end{keywords}

\section{Introduction}
The outflows take place in high-mass star forming regions at very early evolutionary stages~\citep[e.g.][]{birk06,beu07,long11}. They inject large amount of hot gas and kinetic energy into the molecular cloud, causing intense shock waves that dramatically alter the chemistry of the surrounding environment~\citep{dish98}. However, the morphological and dynamical properties of the outflows on smaller scales, as well as their chemical effect to the young high-mass stellar cores are still to be further characterized.

The high-mass star forming region G23.44-0.18 (G23.44 hereafter) was previously detected as a group of strong CH$_3$OH masers~\citep{walsh98} which indicate the presence of young massive stars. It has a trigonometric-parallax distance of 5.88 kpc~\citep{brun09}. In this letter, we report an observation of the massive dusty cores MM1 and MM2 associated with the CH$_3$OH masers, as well as a high-velocity, intense bipolar outflow revealed in CO (2--1). The outflow is causing a shock emission and may be affecting the chemistry in MM2. Section 3 presents the observational results. A further discussion on the outflow is given in Section 4.1 and 4.2. The properties of the dusty cores are discussed in Section 4.3. A summary is given in Section 5.

\section{Observations and Data Reduction}
The observational data is taken from the released SMA data archive\footnote{http://www.cfa.harvard.edu/sma/}. The observation toward G23.44 was carried out in September 2008. The phase tracking center is RA.(J2000)=18$^{\rm h}$34$^{\rm m}$39.25$^{\rm s}$,  Dec.(J2000)=$-8^{\circ}31'36.2''$. The pointing accuracy of the SMA is $\sim3''$. The observation employed 8 antennas in their compact configuration. The correlator has a total bandwidth of 4 GHz, centered at 220 GHz (LSB) and 230 GHz (USB). The on-source integration time is 80 min. The SMA primary beam size (field of view) at this waveband (1.3 mm) is $55''$. The correlator has a channel width of 0.812 MHz (1.1 km s$^{-1}$). The system temperature is between 120 to 150 K. QSO 3c454.3 and Neptune were taken as bandpass and flux calibrators, respectively. QSO 1733-130 and 1911-201 were observed in every 30 and 50 minutes respectively during the observation sequence to track the antenna gains. The interferometer array has a shortest baseline of 17 m, corresponding to an u-v coverage for structures smaller than $20''$.

The calibration and imaging were performed in Miriad$^1$. The absolute flux level has an uncertainty of $\sim10\%$. The continuum map was extracted from the line-free channels in LSB. As the visibility data are INVERTed to the image domain, the synthesized beam size is $4.1''\times3.7''$.

Images of the G23.44 region at other wavelengths were also extracted from several public archives. The Spitzer/IRAC images and the point-source catalogue are taken the database of the GLIMPSE sky survey in the NASA/IPAC Infrared Science Archive\footnote{http://irsa.ipac.caltech.edu/}. We looked for the near-infrared counterparts in the 2MASS point-source catalogue$^2$. The continuum image at JCMT 450 $\micron$ was also obtained to measure the flux densities. They are taken from the Canadian Astronomy Data Center (CADC) repository of the SCUBA Legacy Fundamental Object Catalogue\footnote{http://www4.cadc-ccda.hia-iha.nrc-cnrc.gc.ca}.

\begin{figure}
\centering
\includegraphics[angle=0,width=0.35\textwidth]{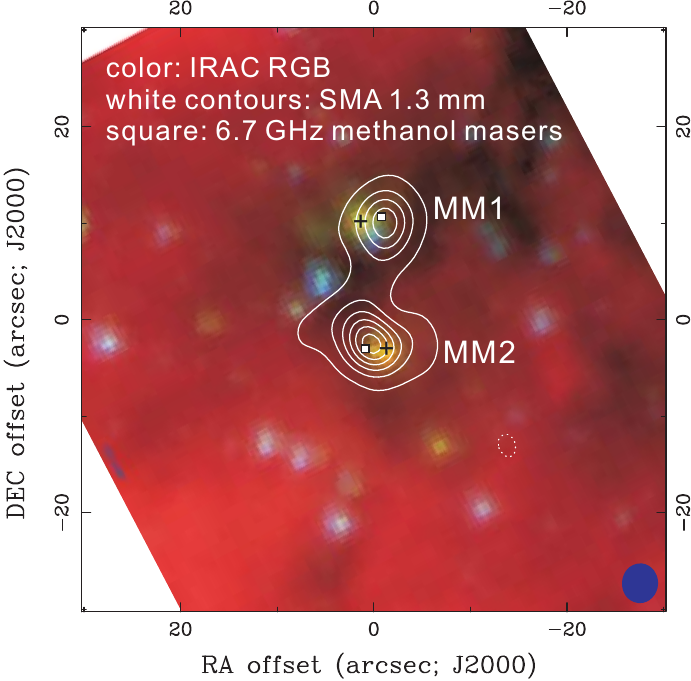} \\
\caption{\small The 1.3 mm continuum emission observed with the SMA overlayed on the IRAC composite image (blue=3.6, green=4.5, and red=8.0 $\micron$) image. The image center is RA.(J2000)=18$^{\rm h}$34$^{\rm m}$39.25$^{\rm s}$ and Dec.(J2000)=$-8^{\circ}31'36.2''$. The contour levels are -5, 5, 15, 25, 35, 45, and 55 $\sigma$ (0.013 Jy beam$^{-1}$). The synthesized beam ($4.1''\times3.7''$) with $PA=7^{\circ}$ from the north to NW. The squares denote the two strongest CH$_3$OH masers. The crosses are the centers of the IRAC sources.}
\end{figure}

\begin{table}
{\small
\centering
\begin{minipage}{80mm}
\caption{Physical parameters of the cores.}
\begin{tabular}{lccl}
\hline\hline
parameter                    &  MM1                        & MM2                & Unit                \\
\hline
$m_J~^a$                        &  $12.315\pm0.049$           & $>15.8$            & mag \\
$m_H~^a$                        &  $11.702\pm0.028$           & $>15.1$            & ... \\
$m_{K_s}~^a$                    &  $11.435\pm0.023$           & $>14.3$            & ... \\
$F(3.6 \micron)^b$             &  $1.42\pm0.28$              & $0.63\pm0.08$      & mJy   \\
$F(4.5 \micron)^b$             &  $25.65\pm2.23$             & $14.48\pm1.96$     & ...   \\
$F(5.8 \micron)^b$             &  $47.84\pm2.22$             & $53.16\pm2.11$     & ...   \\
$F(8.0 \micron)^b$             &  $32.92\pm2.04$             & $40.64\pm2.86$     & ...   \\
$F(450 \micron)^c$             &  $27.3\pm1.5$               & $38.6\pm1.5$       & Jy    \\
F(1.3 mm)                    &  $0.65\pm0.03$              & $1.42\pm0.03$      & ...    \\
$T_{\rm rot}$(CH$_3$CN)      &  $65\pm10$                  & $110\pm10$         & K     \\
$T_{\rm rot}$(CH$_3$OH)      &  $66\pm5$                   & $60\pm6$           & ...     \\
Mass                         &  $120\pm30$                 & $140\pm10$         & $M_{\odot}$  \\
$N({\rm H_2})$               &  $1.9\pm0.2$                & $2.5\pm0.2$        & $10^{23}$ cm$^{-2}$   \\
\hline
\end{tabular}

Note. The flux densities at IR and sub-millimeter wavebands are taken from $^a$2MASS, $^b$Spitzer/IRAC, $^c$JCMT/SCUBA. The remaining data all come from the SMA observation.
\end{minipage}
}
\end{table}

\section{Results}
Figure 1 shows the 1.3 mm continuum emission (white contours) overlayed on the RGB image of the three IRAC bands. It reveals two dust cores aligned from the north to south, denoted as MM1 (northern core) and MM2 (southern core). Deconvoluted with the beam size, MM1 has an mean diameter of $7.3''$ (0.2 pc) while MM2 is more elliptical, with an extent of $9.7''\times6.7''$ ($0.27\times0.19$ pc) and position angle $PA=45^{\circ}$ northeast. The strongest 6.7 GHz CH$_3$OH masers~\citep{walsh98} coincide with the centers of MM1 and MM2. We referred to the MAGPIS 6 cm survey\footnote{http://third.ucllnl.org/gps/index.html} for the potential radio continuum emission from the ionized gas. It turns out that neither MM1 nor MM2 has detectable emission above the sensitivity level of 2.9 mJy, indicating the both cores being prior to forming an Ultra-Compact H{\sc ii}(UC H{\sc ii}) region. The fluxes (or magnitudes) of MM1 and MM2 at near- to far-Infrared are drawn from the 2MASS and IRAC point-source catalogues. The 450 $\micron$ fluxes are measured directly from the JCMT images. The obtained values are shown in Table 1.

\begin{figure*}
\centering
\includegraphics[angle=0,width=0.8\textwidth]{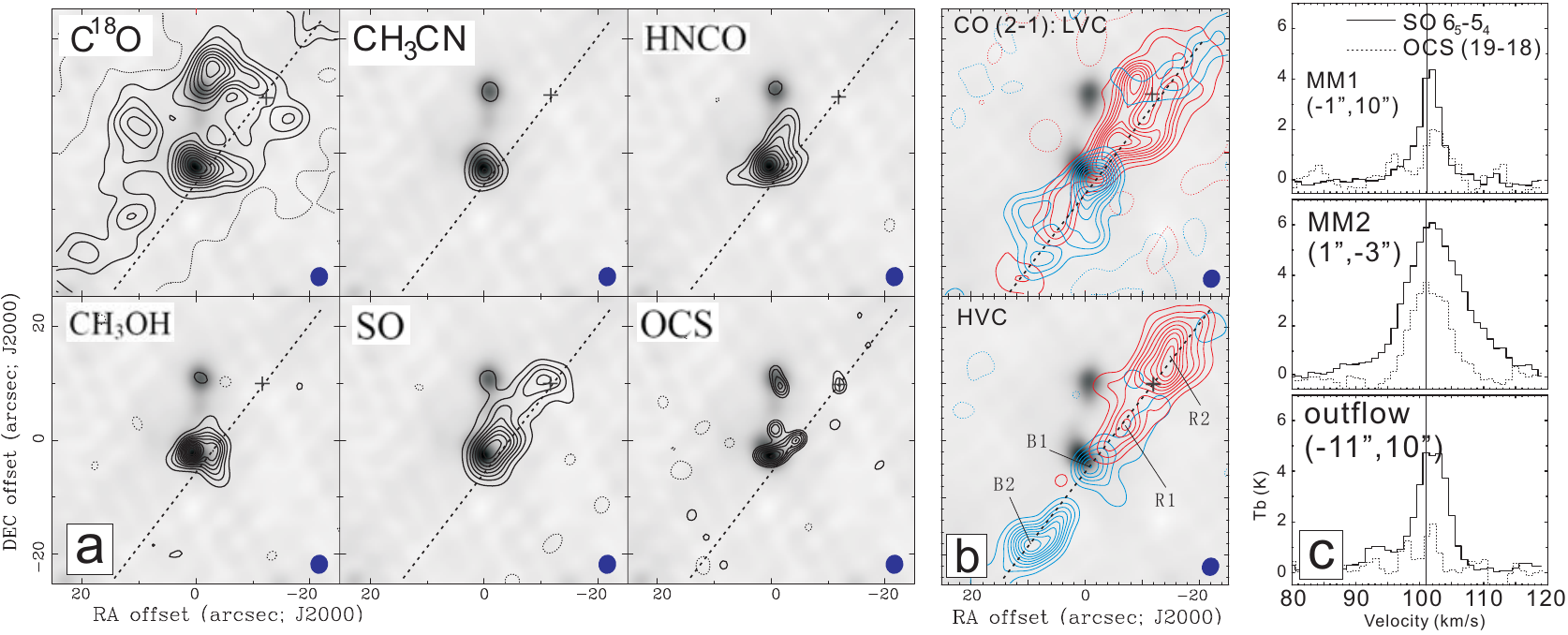} \\
\caption{\small (a) The velocity-integrated images of the molecular lines (contours) overlayed on the 1.3 mm continuum (gray). The integration is from 95 to 110 km s$^{-1}$. For C$^{18}$O (2--1), the contours are -4, 4, 8, 12... 40 $\sigma$ ($1.35$ Jy beam$^{-1}$ km s$^{-1}$). For CH$_3$CN $(12_2-11_2)$, the contours are -4, 4, 8... 24  $\sigma$ ($0.6$ Jy beam$^{-1}$ km s$^{-1}$). For HNCO ($10_{0,10}-9_{0,9}$), the contours are -4, 4, 6, 8... 20 $\sigma$ ($0.34$ Jy beam$^{-1}$ km s$^{-1}$). For CH$_3$OH $(8_0-7_1)$, the contours are -4, 4, 8... 48 $\sigma$ ($0.5$ Jy beam$^{-1}$ km s$^{-1}$). For SO ($6_5-5_4$), the contours are -4, 4, 8, 12... 32 $\sigma$ ($0.6$ Jy beam$^{-1}$ km s$^{-1}$). For OCS (19--18), the contours are -4, 4, 5... 11 $\sigma$ ($0.3$ Jy beam$^{-1}$ km s$^{-1}$). The cross denotes the position of the shocked clump at $(\Delta\alpha=-11'',\Delta\delta=10'')$. (b) The components of the $^{12}$CO (2--1) outflow. The low velocity component (LVC) is integrated in (83,93) and (113,123) km s$^{-1}$ for its blue- and redshifted lobes, respectively, while the high velocity component (HVC) is integrated in (65,75) and (140,170) km s$^{-1}$ for the two lobes. For each lobe, the contours are -10, 10, 20, ..., 90\% of the peak value. The dashed line denotes the axis of the outflow as indicated by the HVC. (c) The beam-averaged SO and OCS spectra at the center of MM1 and MM2, and the shocked clump. The vertical lines denote the systemic velocity of $v_{\rm lsr}=101$ km s$^{-1}$.}
\end{figure*}

The $J=2-1$ transition of $^{12}$CO, $^{13}$CO and C$^{18}$O and a number of molecular species tracing dense gas are detected in our sidebands, including HNCO ($10_{0,10}-9_{0,9}$), SO ($6_5-5_4$), OCS (19--18), CH$_3$CN $12_K-11_K$ ($K=0$ to 6), and four thermally excited CH$_3$OH lines. In the online material, their spectra and physical parameters are provided in Figure S1 and Table S1. The 2MASS $K_s$ and JCMT 450 $\micron$ images are presented in Figure S2 and S3.

Figure 2a shows the integrated images of the C$^{18}$O, CH$_3$CN, HNCO, CH$_3$OH $(8_0-7_1)$, SO, and OCS lines. The C$^{18}$O has a broad distribution, with the emission peaks reasonably coincident with the dust cores. The other species show more compact emissions closely associated with the dust cores. The C$^{18}$O emission has comparable intensities for MM1 and MM2, while the emissions of other dense-gas tracers are much brighter in MM2. Since the dust continuum and C$^{18}$O emission all closely follow the H$_2$ distributions, the intensity contrast suggests that MM2 have higher molecular abundances than MM1, as discussed in Section 4.3.

A strong bipolar outflow is observed in $^{12}$CO (2--1). Figure 2b presents the CO (2--1) image integrated in the four velocity intervals. The outflow can be divided into a pair of low-velocity component (LVC) and high-velocity component (HVC) based on their distinct morphologies. The HVC roughly has $|v|\geq20$ km s$^{-1}$ from the systemic velocity (101 km s$^{-1}$). The bulk of the outflow emission is in the southeast-northwest direction, with a position angle of $PA=40^{\circ}$ northwest. Figure 2c shows the beam-averaged SO and OCS spectra at MM1 and MM2 center, and the shocked region $(\Delta\alpha=-11'',\Delta\delta=10'')$. The $^{13}$CO emission shows a similar feature with $^{12}$CO despite being weak in the HVC. We present the channel maps of $^{12}$CO and $^{13}$CO in Figure S4 and S5, respectively.

\section{discussion}
\subsection{The physical properties of the outflow}
The presence of intense bipolar outflow in G23.44 suggests that the central stars are likely being formed via a disk-mediated accretion~\citep{zhang07}. The HVC is well collimated along the outflow axis, while the LVC, especially its red wing, has a more opened parabolic shape with a decreasing brightness from the edge to the center. The overall geometry of the outflow is reminiscent of the outflow system in the low-mass protostar HH211 \citep{gueth99}. HH211 consists of a high-velocity jet-driven flow and a entrained low-velocity gas shell. The G23.44 outflow may represent a scaled-up version of this system, despite its collimation and shell structure being much less perfect than HH211. However, the size of the outflow in HH211 is only $\sim10^4$ AU, while the G23.44 outflow ($10^5$ AU) has a 10-time larger spatial extent. On such a large scale, the outflow structure might begin to deteriorate due to increased turbulence in its environment.

\begin{table}
{\small
\centering
\begin{minipage}{80mm}
\caption{Abundance of the detected molecules.}
\begin{tabular}{lccc}
\hline\hline
species                        &  $N_{\rm x}$        & $f_{\rm x}$           &  $f_{x,2/1}^a$  \\
(X)                            &  (cm$^{-2}$)        & \quad                 &  \quad \\
\hline
\multicolumn{4}{l}{MM1 $(-1'',10'')$} \\
C$^{18}$O                    &  $3.16\pm0.05(16)$    & $1.7\pm0.1(-7)$         &   $-$ \\
CH$_3$OH                     &  $2.5\pm1.2(15)$      & $1.3\pm0.6(-8)$         &   $-$ \\
CH$_3$CN                     &  $5.2\pm0.6(14)$      & $2.7\pm0.5(-9)$         &   $-$ \\
HNCO                         &  $3.8\pm0.2(14)$      & $2.0\pm0.4(-9)$         &   $-$ \\
SO                           &  $2.5\pm0.2(14)$      & $1.3\pm0.3(-9)$         &   $-$ \\
OCS                          &  $3.7\pm0.2(14)$      & $1.9\pm0.3(-9)$         &   $-$ \\
\hline
\multicolumn{4}{l}{MM2 $(1'',-3'' )$} \\
C$^{18}$O                    &  $7.30\pm0.07(16)$     & $2.9\pm0.1(-7)$        &   $1.7\pm0.2$  \\
CH$_3$OH                     &  $9.8\pm0.8(15)$       & $3.9\pm0.9(-8)$        &   $3.0\pm1.5$  \\
CH$_3$CN                     &  $1.6\pm0.2(15)$       & $6.3\pm0.7(-9)$        &   $2.3\pm0.8$  \\
HNCO                         &  $2.20\pm0.05(15)$     & $8.8\pm1.5(-9)$        &   $4.4\pm1.4$  \\
SO                           &  $1.45\pm0.06(15)$     & $5.8\pm0.9(-9)$        &   $4.5\pm1.4$  \\
OCS                          &  $1.20\pm0.05(15)$     & $4.8\pm0.8(-9)$        &   $2.5\pm0.7$  \\
\hline
\multicolumn{4}{l}{outflow-shock $(-11'',10'')$} \\
C$^{18}$O                   &  $2.4\pm0.07(16)$       & $=2.0(-7)^b$           &  $-$    \\
CH$_3$OH                    &  $2.8\pm0.5(15)$        & $2.3\pm0.5(-8)$        &  $-$    \\
HNCO                        &  $4.1\pm0.2(14)$        & $8.3\pm0.4(-10)$       &  $-$    \\
SO                          &  $3.5\pm0.3(14)$        & $3.1\pm0.3(-9)$        &  $-$    \\
OCS                         &  $8.0\pm0.3(14)$        & $6.6\pm0.3(-9)$        &  $-$    \\
\hline
\end{tabular}

$a.${ The abundance ratio of MM2 to MM1.} \\
$b.${ Assumed as the typical value in ISM. Abundances of the other molecules are then calculated by comparing to $N({\rm C^{18}O})$.}
\end{minipage}
}
\end{table}

The CO outflow, especially the LVC also exhibits a weak emission feature in opposite direction to the major one, i.e. a blueshifted emission in the NW and a redshifted one in the SE. This can be a second flow by chance aligned in the same position angle with the major one. Alternatively, once the outflow direction is close to the plane of the sky and its inclination angle becomes smaller than the opening angle $(\sim30^{\circ})$, its front and back sides will exhibit opposite Doppler shift. This scenario is well demonstrated by~\citet[][Figure 10 therein]{yen10}. We speculate the second case to be more plausible.

Assuming a local thermal equilibrium (LTE) and low optical depth for the $^{12}$CO line-wing emission, the CO column density and the outflow mass can be inferred following the approach of~\citet{gard91}. The excitation temperature of the outflow is speculated to be slightly lower than in the dense core MM2 ($\sim70$ K, Section 4.3). A value of $T_{\rm ex}=50$ K is adopted here. The outflow mass is estimated in each 1 km s$^{-1}$ interval, then the total mass, momentum, and kinetic energy are calculated from $M=\Sigma_v m(v)$, $P=\Sigma_v m(v) v$, and $E_k=\Sigma_v m(v) v^2/2$. In calculation, we assume an inclination angle of $\theta=20^\circ$ to correct the velocity projection. The outflow turns out to have $M\simeq5.5~M_{\odot}$, $P\simeq360~M_{\odot}$ km s$^{-1}$, $E_k\simeq9\times10^{46}$ erg.

Each of the blue and red wings in the HVC is resolved into two major clumps (Figure 2b, lower panel) denoted B1, B2 and R1, R2. These clumps in the outflow may indicate an episodic mass ejection~\citep[e.g.][]{qiu09}. The average central distance is $24''$ (0.68 pc) for the pair of B1-R1, and $5.0''$ (0.14 pc) for B2-R2. The two pairs both show a reasonable geometrical symmetry. However, the clumps are not exactly aligned across the MM2 center, instead with an offset to its west. This may indicate that the driven agent, or the stellar disk system is also displaced from the MM2 center. The dynamical age of the outflow $t_{\rm dyn}$ is estimated by dividing its spatial extent with the typical velocity ($v=40/\sin\theta$ km s$^{-1}$). That yields $t_{\rm dyn}=6.1\times10^3$ years for the B1-R1 pair and $t_{\rm dyn}=1.3\times10^3$ years for B2-R2. Adopting the first value, we calculate the mass ejection rate to be $\dot{M}=M/t_{\rm dyn}\simeq1.0\times10^{-3} M_{\odot}$ yr$^{-1}$. One can see that the outflow is young and performing an intense mass ejection.

\subsection{Chemical enhancement in the outflow and shock}
In both low- and high-mass young stellar objects the outflows have led to a rich chemistry in CH$_3$OH, HNCO, and sulphides~\citep[][etc.]{blake87,tak03,jog04,rod10}. In G23.44, the SO and OCS emission shows a noticeable emission clump at offset=$(-11'',10'')$ (Figure 2a).  The clump is almost perfectly aligned with the HVC axis of the CO (2--1), and reasonably coincides with the peak R2, despite being closer to MM2 center. This strongly suggests that the clump traces a shock within the jet/outflow. The clump is also weakly detected in HNCO and CH$_3$OH (spectra shown in Figure S1). Their column densities are estimated using Equation (1) and (2) in Section 4.3. In calculation a temperature of $T_{\rm ex}=50$ K [same with that of CO (2--1)] is adopted, and the C$^{18}$O abundance is taken to be a constant of $2\times10^{-7}$ as the typical ISM abundance which is close to $f_{\rm C^{18}O}$ measured in MM1 and MM2. The abundances of other molecules are then inferred from their intensity ratio relative to C$^{18}$O (2--1). The abundances of the shocked clump are listed in Table 2.

In the low-mass protostars, the outflows often lead to a rich chemistry. In a few cases, the column density of HNCO and S-bearing molecules can been enhanced for orders of magnitude \citep[e.g. L1157 and L1448,][]{rod10,taf10}. In comparison, the shocked region at $(-11'',10'')$ shows moderate abundances which are comparable or even slightly lower than in MM2. The enhancement level can be limited by the temperature and density in the shocked region, as well as the efficiency of the previous grain-surface chemistry. In addition, since the outflow has a short $t_{\rm dyn}$, the shock chemistry may still be undergoing and have not yet attained its maximum level.

The blue lobe of CO $(2-1)$ outflow are devoid of the SO and OCS emissions. As a possible explanation, the S-bearing species (e.g. H$_2$S and CS as their progenitors) would be rather tightly bound to the dust grains~\citep{blake87}, and and a threshold shock velocity may be required for their desorption. Higher resolution and primary shock products like SiO and H$_2$S may better reveal the shock chemistry.

\begin{figure}
\centering
\includegraphics[angle=0,width=0.4\textwidth]{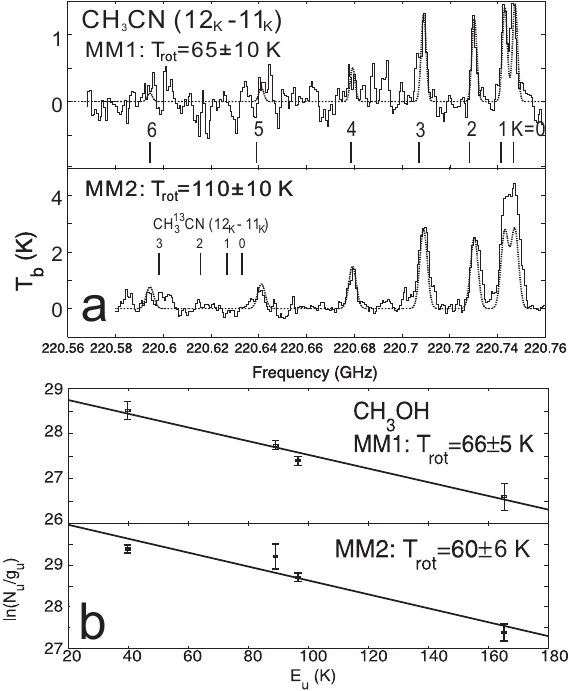} \\
\caption{\small (a) The solid line is the CH$_3$CN ($12_K-11_K$) towards the center of MM1 (top panel) and MM2 (bottom panel). The dashed line is the spectrum of the best fit from the LTE gas model. (b) The rotation diagram of the CH$_3$OH lines. The filled circles are for the observed transitions. The vertical bars indicate the $3\sigma$ errors. The linear least-squares fit is shown as the solid line.}
\end{figure}

\subsection{The dense core properties}
The dust cores MM1 and MM2 coincide with dark patches on the extended 8 $\micron$ emission (Figure 1). The two cores are slightly connected with each other, suggesting that they might be the fragmentation products from the same natal cloud.

The CH$_3$CN ($12_K-11_K$) lines provide a temperature diagnosis for the cores (Figure 3a). We take an LTE radiative transfer model with a uniform rotation temperature and gas density~\citep{chen06,zhang07} to reproduce the observed CH$_3$CN spectra at the dust-core center. We found the best fit at $T_{\rm rot}=65$ K for MM1 and $T_{\rm rot}=110$ K for MM2. The modeled spectra of $K=0$ and 1 towards MM2 is much lower than the observed line profile while the $K>1$ lines are all well fitted. The excess emission in $K=0$ and 1 lines may largely arise from the cold outer-envelope gas.

The rotation temperature is also evaluated from the thermally excited CH$_3$OH lines at the core center using the rotation diagram method~\citep{gold99,liu01,biss07}. The column density of the upper level $N_{\rm u}$ is related to the line intensity in the form
\begin{equation}
\qquad \qquad \frac{N_{\rm u}}{g_{\rm u}}=\frac{3k}{8\pi^3\nu S\mu^2}\int T_{\rm b} \, {\rm d}v
\end{equation}
where $g_{\rm u}$ is the upper-level degeneracy. $S$ is the line strength, $\mu$ is the dipole moment. $N_{\rm u}$ depends on the total column density $N_{\rm T}$ in the form
\begin{equation}
\qquad \qquad \ln(\frac{N_{\rm u}}{g_{\rm u}})=\ln(\frac{N_{\rm T}}{Q_{\rm rot}})-\frac{E_{\rm u}}{T_{\rm rot}}
\end{equation}
where the partition function $Q_{\rm rot}$ can be adopted from~\citet{blake87}. $T_{\rm rot}$ is then estimated from a least-square fit (Figure 3b). The derived values are listed in Table 1.

The $T_{\rm rot}$ derived from CH$_3$OH and CH$_3$CN are remarkably similar for MM1. For MM2, however, the $T_{\rm rot}$ of CH$_3$OH is lower than that of CH$_3$CN. As a probable reason, the CH$_3$OH lines generally have lower $E_{\rm u}$ than CH$_3$CN ($12_K-11_K$), then the cooler outer-part gas would contribute more significantly to the CH$_3$OH lines than to the CH$_3$CN (in particular $K=2$ to 6). In addition the CH$_3$OH can be easily released from the dust grains by the shocks while the CH$_3$CN production requires subsequent gas-phase reactions and stellar heating~\citep{dish98}. CH$_3$CN may therefore be distributed closer to the MM2 center (as seen in Figure 2a), and showing a higher $T_{\rm rot}$.

For the other species, an independent temperature estimation is currently unavailable. We adopt $T_{\rm ex}=60$ K for MM1 and $T_{\rm ex}=70$ K for MM2 assuming an LTE excitation, the column density of the molecules are calculated from their spectral lines towards the dust-core center, using equation (1) and (2).

Assuming optically thin 1.3 mm dust continuum emission, the total dust-and-gas mass can be estimated using $S_{\nu}=M_{\rm core}\kappa_{\nu}B_{\nu}(T_{\rm d})/g D^2$. In the formula $S_{\nu}$ is the total flux density at 1.3 mm. $g=100$ is the dust-gas mass ratio. $B_{\nu}(T_{\rm d})=(2h\nu^3/c^2)/[\exp(T_0/T_{\rm d})-1]$ is the Planck function ($T_0=h\nu/k$). The dust opacity is set to be $\kappa_{\nu}=\kappa_{\rm 230 GHz}=0.9$ cm$^2$ g$^{-1}$ of MRN dust grains with thin ice mantles at $n_{\rm H_2}=10^6$ cm$^{-3}$~\citep{ossen94}. In the expression of $S_{\nu}$, using the beam-averaged flux density towards the core center, and dividing the obtained mass with $m_{\rm H_2}$, one can also calculate the peak column density N(H$_2$) at MM1 and MM2. The derived values are presented in Table 1. MM2 has a comparable but slightly higher N(H$_2$) than MM1.

The molecular abundances are estimated using $f_{\rm x}=N(X)/N({\rm H_2})$. The $f_{\rm x}$ and its ratio between MM2 and MM1 are listed in Table 2. For the high-density tracers, the abundances in MM2 are generally 3 to 4 times higher than those in MM1. The SO shows the highest abundance ratio of 4.5. We note that the temperature adopted for MM2 (70 K) is conservative. A higher $T_{\rm ex}$ would yield a lower N(H$_2$) and higher $N(X)$ for most species, hence leading to even higher abundances. For instance, adopting the value of $T_{\rm rot}({\rm CH_3CN})=110$ K as the $T_{\rm ex}$ for MM2, we will have $N({\rm HNCO})=6.7\times10^{15}$ cm$^{2}$ that is 3 times larger than the current value.

The molecular distribution appears to be influenced by the outflow. Besides the coherent elongation of SO and OCS with the HVC of $^{12}$CO emission (Figure 2b), the CH$_3$OH and HNCO emissions are also largely distributed to the west of MM2. In addition, all the molecular lines have broader line widths in MM2 than in MM1, e.g. The SO line has $\Delta v_{\rm FWHM}=11.3$ in MM2 and 3.4 km s$^{-1}$ in MM1. The line widths of MM2 can be broadened by a kinetic energy input from the outflow. Considering its influence to the molecular line profiles, the outflow may also contribute to the higher abundances in MM2. Specifically, there would be underlying shocks which are induced as the outflow is interacting with the envelope and/or infalling gas~\citep{arce04,wu09}.

At the near- to mid-IR and sub-millimeter wavebands, the two cores have similar flux densities (Table 1). At IRAC bands MM2 is brighter than MM1 at 5.8 and 8 $\micron$ but weaker at 3.6 and 4.5 $\micron$, while at 2MASS $J$, $H$, $K_s$ bands, only MM1 is significantly detected (Figure S2). Having a spectral energy distribution (SED) obviously skewed to the shorter wavelengths ($\lambda<4.5~\micron$), MM1 may have a more evolved stage for its stellar disk system~\citep{rob06}. However, since the outflow in MM2 is close to the plane of the sky, it is also possible that a disk/toroid structure lies edge-on in MM2 thus would largely absorb the short-wave emissions. A higher resolution will be helpful in revealing the morphology of the stellar system and underlying shocks in MM2.

\section{summary}
We present a multi-wavelength study towards the young high-mass star forming region G23.44-0.18. Two dusty cores, MM1 and MM2 are observed in 1.3 mm continuum emission and the molecular lines of CH$_3$OH, CH$_3$CN, HNCO, SO, and OCS. The both cores have a pre-UC H{\sc ii} evolutionary stage. A strong bipolar outflow is arising from MM2 as revealed by CO (2--1). The outflow consists of a collimated, bi-polar high-velocity component and a more extended, parabolic flow at lower velocities. A clump of shocked gas is observed in SO $(6_5-5_4)$ and OCS $(19-18)$ lines and also detected in HNCO and CH$_3$OH. The outflow shows a high momentum and mass-loss rate, and a short timescale of $t_{\rm dyn}\sim6\times10^3$ years.

In MM2 the high-density tracers CH$_3$OH, CH$_3$CN, HNCO, SO, and OCS show abundances of 3 to 4 times higher than those in MM1. To explain this difference, we suggest a combined effect of the stellar heating and underlying shock interactions in MM2. A further study with higher excited lines and improved resolution can be performed to probe the chemistry, possible disk-jet system and mass accretion close to the MM2 center.

\section*{Acknowledgment}
We are grateful to the SMA observers and the SMA data archive. We would thank the anonymous referee for the useful comments that helped to improve the presentation and interpretation. This work is supported by grants of No.10733033, 10873019, 10973003, 2009CB24901, and the Doctoral Candidate Innovation Research Support Program (kjdb201001-1) from \textit{Science \& Technology Review}.

\clearpage

\setcounter{figure}{0}
\setcounter{table}{0}
\renewcommand\thefigure{S\arabic{figure}}
\renewcommand\thetable{S\arabic{table}}

\begin{table*}
\centering
{\small
\caption{Parameters of the molecular lines.}
\begin{tabular}{lllccc}
\hline\hline
Frequency    &     Species            & \quad                   &  $V_{\rm lsr}$    &  $T_{\rm b}^{\rm peak}$  &  $\Delta V_{\rm FWHM}$    \\
(GHz)        &     (X)                & Transition              &  (km s$^{-1}$)    &  (K)                     &  (km s$^{-1}$)            \\
\hline
\multicolumn{6}{l}{MM1 $(-1'',10'')$} \\
219.5603...  &      C$^{18}$O         & $2-1$                   &  102.2            &  $11\pm0.4$          &  $3.2\pm0.5$      \\
219.7982...  &      HNCO              & $10_{0,10}-9_{0,9}$     &  101.4            &  $1.8\pm0.1$         &  $3.6\pm0.4$      \\
219.9494...  &      SO                & $6_5-5_4$               &  101.3            &  $4.2\pm0.1$         &  $3.4\pm0.4$      \\
220.0784...  &      CH$_3$OH          & $8_{0,8}-7_{1,6}$       &  101.5            &  $1.0\pm0.3$         &  $4.2\pm0.6$      \\
220.7302...  &      CH$_3$CN          & $12_2-11_2$             &  101.6            &  $1.3\pm0.2$         &  $5.5\pm0.8$      \\
229.7588...  &      CH$_3$OH          & $8_{-1}-7_0$            &  101.8            &  $3.9\pm0.1$         &  $3.5\pm0.5$      \\
230.0270...  &      CH$_3$OH          & $3_{-2,2}-4_{-1,4}$     &  102.1            &  $1.3\pm0.3$         &  $2.9\pm0.6$      \\
231.0609...  &      OCS               & $19-18$                 &  102.1            &  $1.9\pm0.2$         &  $4.5\pm0.4$      \\
231.2810...  &      CH$_3$OH          & $10_{2,9}-9_{3,6}$      &  101.3            &  $0.6\pm0.2$         &  $2.3\pm0.4$      \\
\hline
\multicolumn{6}{l}{MM2 $(1'',-3'')$} \\
219.5603...  &      C$^{18}$O         & $2-1$                   &  99.8             &  $6.4\pm0.5$         &  $10.5\pm1.5$     \\
219.7982...  &      HNCO              & $10_{0,10}-9_{0,9}$     &  100.6            &  $3.5\pm0.1$         &  $8.8\pm0.6$      \\
219.9494...  &      SO                & $6_5-5_4$               &  101.5            &  $6.0\pm0.1$         &  $11.3\pm0.7$     \\
220.0784...  &      CH$_3$OH          & $8_{0,8}-7_{1,6}$       &  100.5            &  $3.5\pm0.3$         &  $6.7\pm0.8$      \\
220.7302...  &      CH$_3$CN          & $12_3-11_3$             &  101.7            &  $2.8\pm0.1$         &  $6.6\pm0.7$      \\
229.7588...  &      CH$_3$OH          & $8_{-1}-7_0$            &  102.1            &  $15.0\pm0.1$        &  $8.5\pm0.8$      \\
230.0270...  &      CH$_3$OH          & $3_{-2,2}-4_{-1,4}$     &  102.1            &  $2.0\pm0.2$         &  $4.3\pm0.5$      \\
231.0609...  &      OCS               & $19-18$                 &  102.1            &  $3.2\pm0.1$         &  $6.7\pm0.4$      \\
231.2810...  &      CH$_3$OH          & $10_{2,9}-9_{3,6}$      &  102.1            &  $1.4\pm0.3$         &  $4.3\pm0.3$      \\
\hline
\multicolumn{6}{l}{Shock $(-11'',10'')$} \\
219.5603...  &      C$^{18}$O         & $2-1$                   &  102.1            &  $8.0\pm0.3$         &  $2.7\pm0.4$      \\
219.7982...  &      HNCO              & $10_{0,10}-9_{0,9}$     &  96.3             &  $0.8\pm0.2$         &  $1.3\pm0.4$      \\
219.9494...  &      SO                & $6_5-5_4$               &  102.3            &  $4.9\pm0.3$         &  $5.6\pm0.5$      \\
220.0784...  &      CH$_3$OH          & $8_{0,8}-7_{1,6}$       &  102.1            &  $0.9\pm0.2$         &  $3.7\pm0.6$      \\
220.7302...  &      CH$_3$CN          & $12_2-11_2$             &  $-$              &  $-$                 &  $-$              \\
229.7588...  &      CH$_3$OH          & $8_{-1}-7_0$            &  103.2            &  $8.0\pm0.2$         &  $4.2\pm0.4$      \\
230.0270...  &      CH$_3$OH          & $3_{-2,2}-4_{-1,4}$     &  $-$              &  $-$                 &  $-$              \\
231.0609...  &      OCS               & $19-18$                 &  102.1            &  $2.0\pm0.2$         &  $3.3\pm0.5$      \\
231.2810...  &      CH$_3$OH          & $10_{2,9}-9_{3,6}$      &  $-$              &  $-$                 &  $-$              \\
\hline
\end{tabular}
}
\end{table*}

\begin{figure*}
\centering
\includegraphics[angle=0,width=0.6\textwidth]{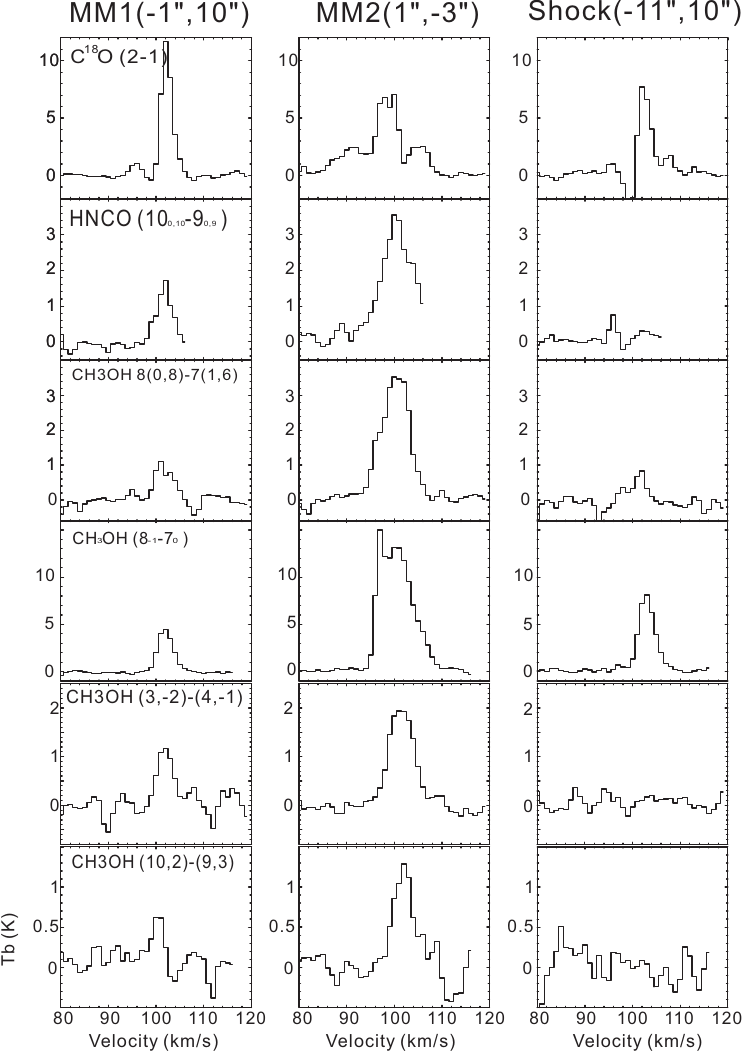} \\
\caption{The molecular lines detected in the observational sidebands at the three positions.}
\end{figure*}

\begin{figure*}
\centering
\includegraphics[angle=0,width=0.4\textwidth]{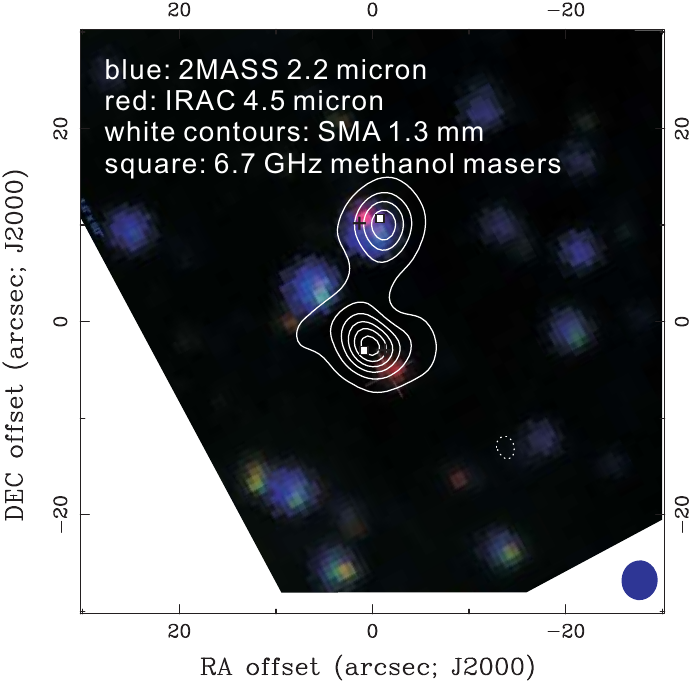} \\
\caption{The near to mid-IR RGB-coded image of G23.44. The blue is the 2MASS $K_s$, the green and red colors are IRAC 3.6 and 4.5 micron, respectively. The image center is RA.(J2000)=18$^{\rm h}$34$^{\rm m}$39.25$^{\rm s}$ and Dec.(J2000)=$-8^{\circ}31'36.2''$. The contour levels are -5, 5, 15, 25, 35, 45, and 55 $\sigma$ (0.013 Jy beam$^{-1}$). The synthesized beam ($4.1''\times3.7''$) with $PA=7^{\circ}$ to the west. The squares are the two strongest CH$_3$OH masers. The crosses are the centers of the IRAC sources. }
\end{figure*}

\begin{figure*}
\centering
\includegraphics[angle=0,width=0.4\textwidth]{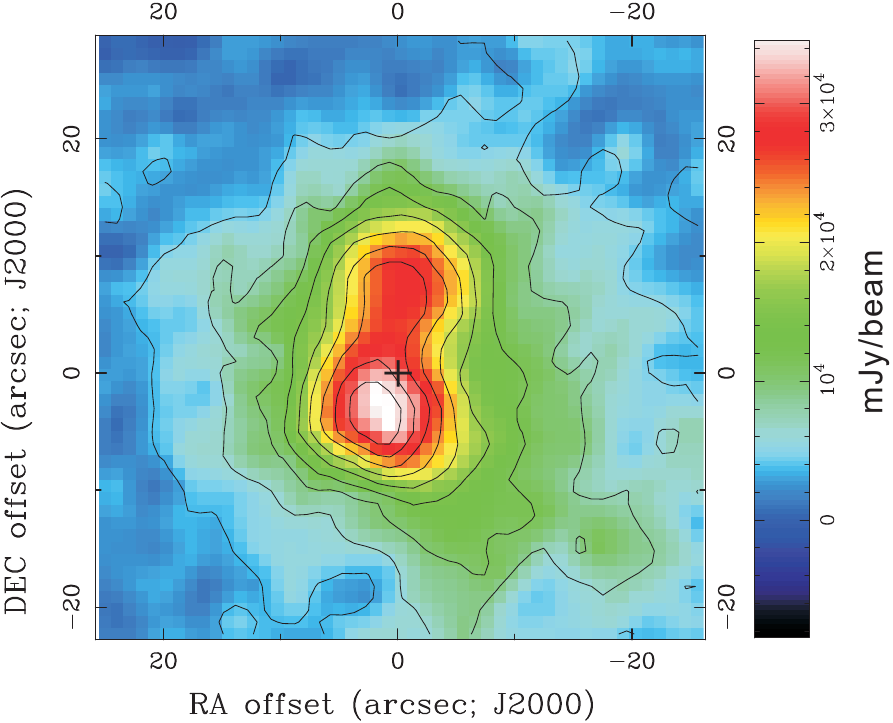} \\
\caption{The SCUBA 450 $\micron$ image of G23.44. Contours are 10 to 90 per cent of the peak brightness (34.5 Jy beam$^{-1}$). The beam size is $7.5''$. The cross denotes the center position RA.(J2000)=18$^{\rm h}$34$^{\rm m}$39.25$^{\rm s}$ and Dec.(J2000)=$-8^{\circ}31'36.2''$. }
\end{figure*}

\clearpage
\begin{figure*}
\centering
\includegraphics[angle=0,width=0.8\textwidth]{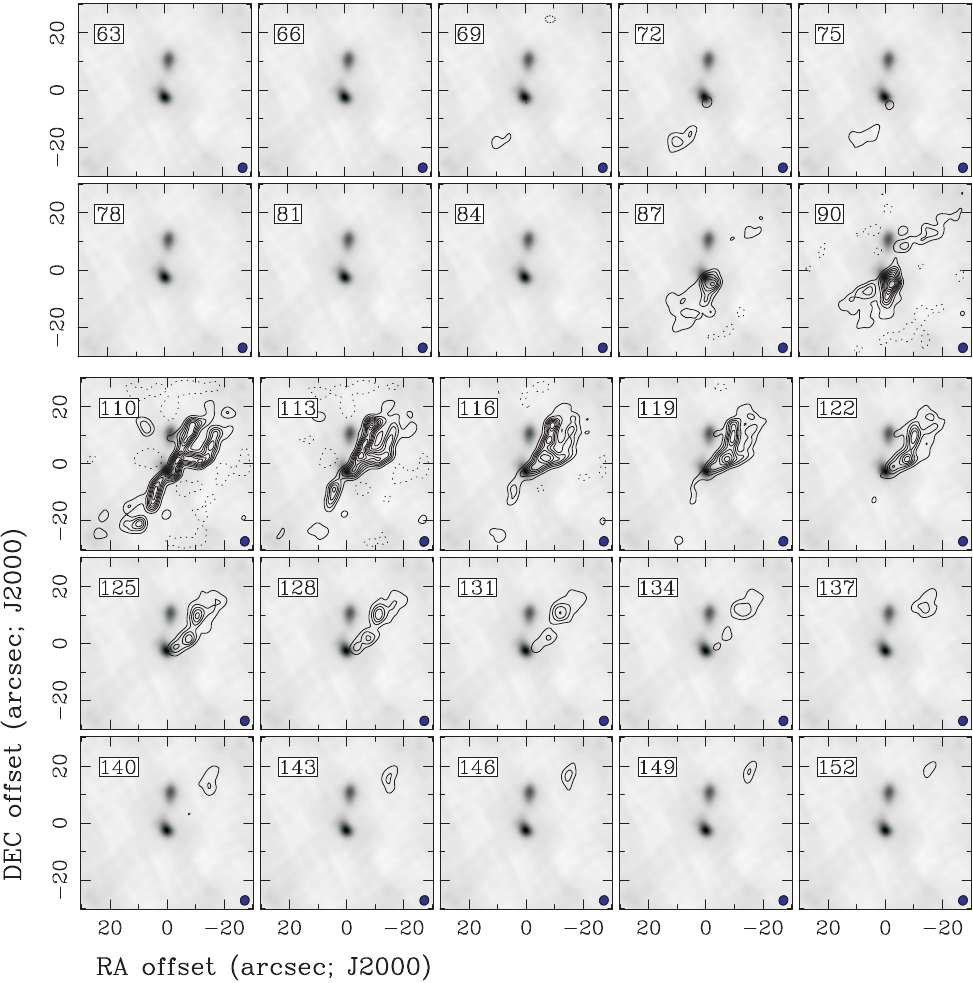} \\
\caption{The $^{12}$CO (2--1) channel images. The contours are -10, 20 to 90 percent of the peak intensity, which is 45 Jy beam$^{-1}$ for the velocity range of [63,90] km s$^{-1}$ and 18.9 Jy beam$^{-1}$ for v=[110,152] km s$^{-1}$. Gray is the SMA 1.3 mm continuum emission.}
\end{figure*}

\begin{figure*}
\centering
\includegraphics[angle=0,width=0.8\textwidth]{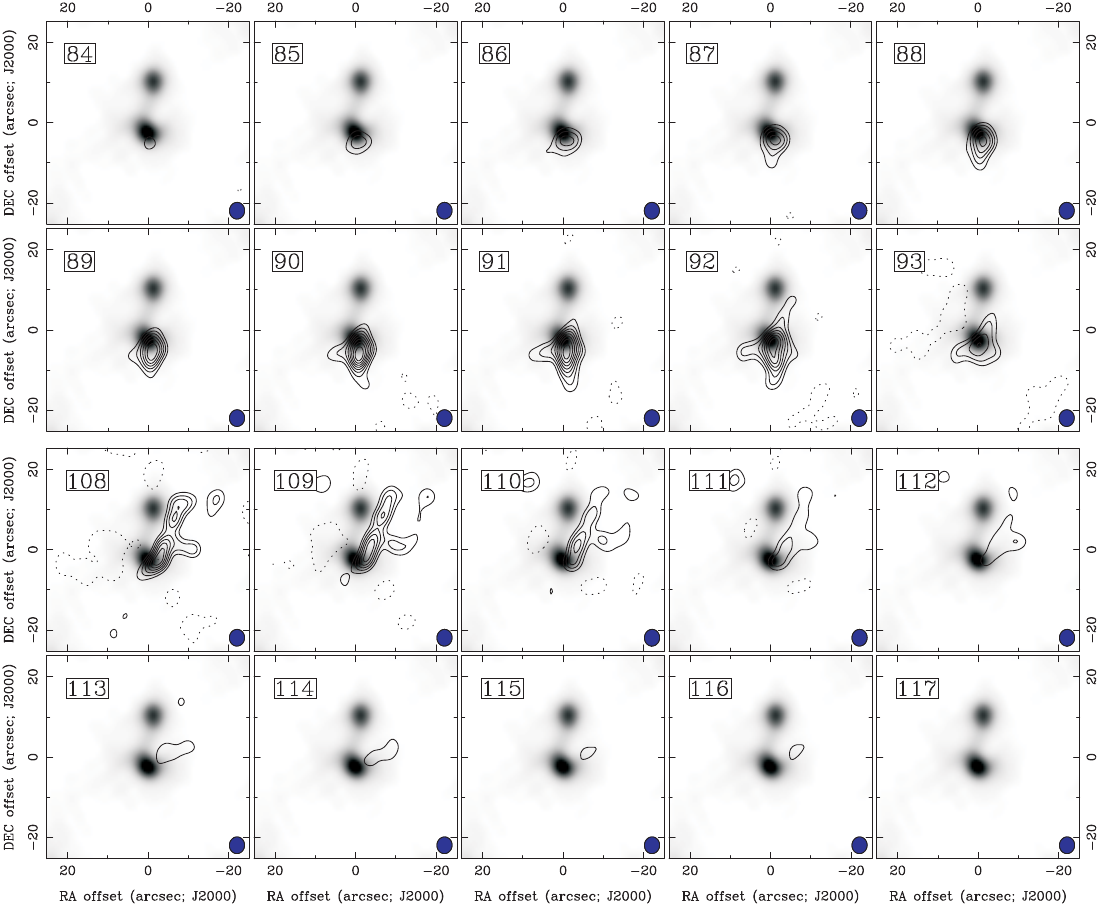} \\
\caption{The $^{13}$CO (2--1) channel images. The contours are -10, 20 to 90 percent of the peak intensity which is 12.2 Jy beam$^{-1}$ for the velocity range of [84,93] km s$^{-1}$ and 8.8 Jy beam$^{-1}$ for v=[108,117] km s$^{-1}$. }
\end{figure*}

\end{document}